\documentclass[prl,twocolumn,amssymb,amsmath,nofootinbib,superscriptaddress]{revtex4-1}

\usepackage{hyperref}
\usepackage{color}
\usepackage{graphicx,psfrag, subfigure}
\usepackage{bbold}

\makeatletter
\newsavebox\myboxA
\newsavebox\myboxB
\newlength\mylenA

\newcommand*\xoverline[2][0.75]{%
    \sbox{\myboxA}{$\m@th#2$}%
    \setbox\myboxB\null
    \ht\myboxB=\ht\myboxA%
    \dp\myboxB=\dp\myboxA%
    \wd\myboxB=#1\wd\myboxA
    \sbox\myboxB{$\m@th\overline{\copy\myboxB}$}
    \setlength\mylenA{\the\wd\myboxA}
    \addtolength\mylenA{-\the\wd\myboxB}%
    \ifdim\wd\myboxB<\wd\myboxA%
       \rlap{\hskip 0.5\mylenA\usebox\myboxB}{\usebox\myboxA}%
    \else
        \hskip -0.5\mylenA\rlap{\usebox\myboxA}{\hskip 0.5\mylenA\usebox\myboxB}%
    \fi}
\makeatother

\begin{document}

\title{A New Angle on Chaotic Inflation}

\author{Thomas C.~Bachlechner} \affiliation{Department of Physics, Cornell University, Ithaca, New York, 14853, USA}

\author{Mafalda Dias}  \affiliation{Astronomy Centre, University of Sussex, Falmer, Brighton, BN1 9QH, UK}

\author{Jonathan Frazer} \affiliation{Department of Theoretical Physics, University of the Basque Country UPV/EHU, 48040 Bilbao, Spain}

\author{Liam McAllister}  \affiliation{Department of Physics, Cornell University, Ithaca, New York, 14853, USA}

\vskip 4pt

\begin{abstract}
N-flation is a  radiatively stable scenario for chaotic inflation in which  the  displacements of $N \gg 1$ axions with decay constants $f_1 \le \ldots \le f_N < M_P$ lead to a super-Planckian  effective displacement  equal to the Pythagorean sum $f_{\rm{Py}}$ of the $f_i$.
We show  that mixing  in the axion kinetic term  generically leads to  the phenomenon of kinetic alignment, allowing for effective displacements as large as $\sqrt{N} f_{N} \ge f_{\rm{Py}}$,
even
if $f_1, \ldots, f_{N-1}$ are arbitrarily small.
At the level of kinematics, the necessary alignment
occurs with very high probability,  because of eigenvector delocalization.  We present conditions under which inflation can take place along an aligned direction.
Our construction  sharply reduces the challenge of realizing N-flation  in string theory.

\end{abstract}

\maketitle

\section{Introduction}

Inflationary scenarios producing detectable primordial gravitational waves are  extraordinarily sensitive to Planck-scale physics, motivating the understanding of these models in
string theory.
The recent observation of B-mode polarization at degree angular scales by the BICEP2  collaboration \cite{Ade:2014xna} provides  the prospect of direct experimental study of  large-field inflation, if the
signal is established as primordial in origin.

Among the best-motivated scenarios for  large-field inflation in string theory are axion inflation models,  including  string-theoretic variants of natural inflation \cite{Freese:1990rb}, in which
shift symmetries protect the inflaton potential (for a recent review, see \cite{Pajer:2013fsa}).  In the effective  field theory description,
axionic shift symmetries with large periodicities, i.e.~with decay constants $f \gg M_P$, can ensure radiative stability  of large-field inflation, but whether such symmetries  admit completions in quantum gravity is a delicate question that requires knowledge of the ultraviolet theory. By
embedding axion inflation in  string theory  one can address this problem through  well-defined computations.

A general finding about axions in presently-understood string vacua is that the decay constants $f$  are small, $f \ll M_P$,  in all regions  in which the perturbative and nonperturbative corrections to the effective action are under parametric control \cite{Banks:2003sx}.   At the same time, axions are very numerous, with ${\cal O}(10^2)-{\cal O}(10^3)$ independent axions appearing in typical compactifications.
To achieve large-field inflation in string theory,  one could therefore consider  a collective excitation of $N \gg 1$ axions $\phi_i$, $i=1,\ldots, N$, with effective displacement $\Delta\Phi$
larger than the displacements $\Delta\phi_i$ of the individual fields.  This proposal, known as {\it{N-flation}} \cite{Dimopoulos:2005ac},  builds on the idea of assisted inflation \cite{Liddle:1998jc}.

If the fields $\phi_i$ are canonically-normalized axions with periodicity $2\pi f_i$
then the diameter of the field space is
\begin{equation}
 {\rm{Diam}} = 2\pi \sqrt{\sum f_i^2} \equiv 2\pi f_{{\rm{Py}}}\, ,
\end{equation} where $f_{{\rm{Py}}}$ denotes the Pythagorean sum of the $f_i$.
In an inflationary model involving small displacements  of each axion around the minimum of the potential,  so that a quadratic approximation to the potential remains valid, the maximum collective displacement is $\Delta\Phi \lesssim c_D\cdot{\rm{Diam}} \approx f_{{\rm{Py}}}$  for some constant $c_D \sim {\cal O}(0.1)$.

Finding  string compactifications containing $N \gg 1$ axions with $f_N\ge\ldots \ge f_1 \gtrsim 0.1 M_P$
appears difficult.
In this Letter we show that given specific well-motivated assumptions about the axion kinetic terms, one can achieve a large effective displacement $\Delta \Phi \approx \sqrt{N} f_N$  even if $f_1,\ldots,f_{N-1}$ are very small.

\begin{figure}
  \centering
  \includegraphics[width=.3\textwidth]{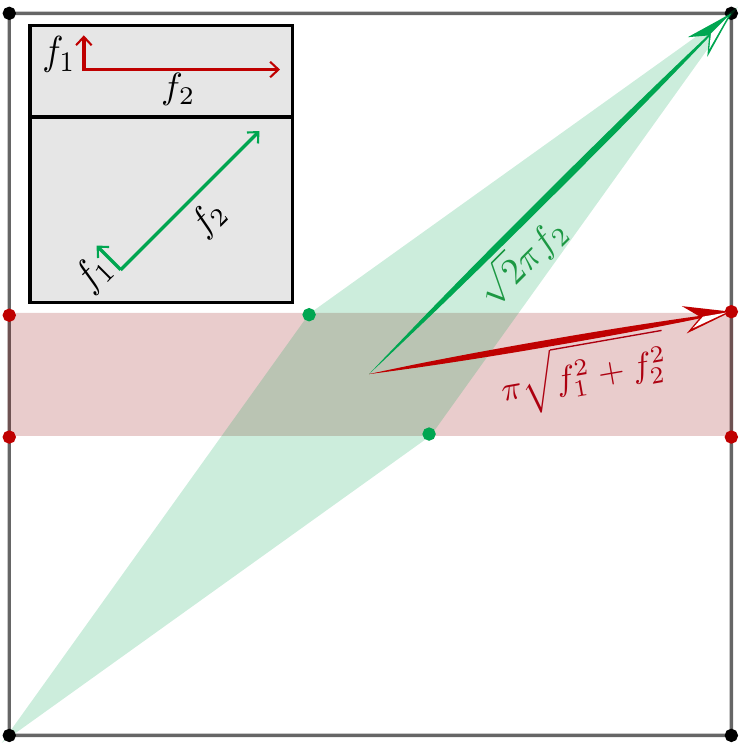}
  \caption{\small N-flation with and without alignment, for $N=2$.  The outer box shows a fundamental domain in the dimensionless $\theta$ coordinates with period $2\pi$.  The shaded  rhombus and rectangle depict the physical  size of the fundamental domain  with and without alignment, respectively.  The  green and red vectors show the semidiameters in the aligned and unaligned cases.  The inset is not to scale.}\label{fig:illustration}
\end{figure}
\section{The Action for $N$ Axions}

Consider a field theory containing $N$ axion fields $\theta_i$ corresponding to $N$ independent
shift symmetries $\theta_i \to \theta_i + c_i$.
We take the symmetries to be broken nonperturbatively by the potential
\begin{equation}
V(\theta_1,\ldots \theta_N) = \sum_i \Lambda_i^4 \bigl[1-{\rm{cos}}(\theta_i)\bigr] + \ldots\  \label{vcos}
\end{equation} to discrete shifts $\theta_i \to \theta_i + 2\pi$, where $\Lambda_i$ are dynamically-generated scales, and the ellipses indicate  terms at higher orders in the instanton expansion.
The potential (\ref{vcos}) breaks the $GL(N,\mathbb{R})$  symmetry of the perturbative Lagrangian:  the periodic identifications define a lattice in the field space.   Without loss of generality, we have taken the periodicities of the dimensionless fields $\theta_i$ to be $2\pi$, so that a fundamental domain is a hypercube of side length $2\pi$ in $\mathbb{R}^N$.  We refer to the corresponding basis,  which is simply the usual Cartesian basis of $\mathbb{R}^N$, as the {\it{lattice basis}}.

Including the kinetic term,  the Lagrangian takes the form
\begin{equation}
{\cal{L}} = \frac{1}{2} K_{ij}\partial\theta^i \partial\theta^j  - \sum_i \Lambda_i^4 \bigl[1-{\rm{cos}}(\theta_i)\bigr]\, ,  \label{basicL}
\end{equation} where $K_{ij}$ has mass dimension two.
We will refer to the dimensionless field space parameterized by the $\theta_i$ as ${\cal M}_{\theta}$, and the physical field space  with metric $K_{ij}$ as ${\cal M}_{K}$.
We refer to the basis in which $K_{ij}$ is diagonal as the {\it{kinetic basis}}.
There is no reason to expect that $K_{ij}$ should be diagonal in the lattice basis: instead, the lattice basis and the kinetic basis will typically be related by a nontrivial rotation.\footnote{Although our considerations are general, a concrete example may be helpful.  In flux compactifications of type IIB  string theory, the kinetic terms for the axions  appearing in K\"ahler moduli multiplets are determined by the intersection numbers, while a nonperturbative potential arises from Euclidean D-branes.  The four-cycles associated to the leading instanton terms are generally nontrivial linear combinations of the coordinates in which the K\"ahler metric is diagonal, so that the lattice  basis differs from the kinetic basis.}

In a  model with a single axion $\theta$,  the axion decay constant $f$ can be defined by changing coordinates
to write the Lagrangian in the form
${\cal{L}} = \frac{1}{2} (\partial\phi)^2   -  \Lambda^4 [1-{\rm{cos}}(\phi/f)]$.
The periodicity of the canonically-normalized field $\phi$ is then $2\pi f$.
In a model with $N$ axions, but for which the lattice basis  is  proportional to the kinetic basis (a non-generic circumstance),
the
Lagrangian can be put in the form
\begin{equation}
{\cal{L}} = \frac{1}{2} (\partial\phi_i)^2 - \sum_i \Lambda_i^4 \bigl[1-{\rm{cos}}(\phi_i/f_i)\bigr]  \ , \label{nfieldssimple}
\end{equation} where $f_i^2$ is the $i$th eigenvalue of $K_{ij}$.
But in a general model with $N$ axions, where the lattice and kinetic bases are not proportional,  the Lagrangian cannot be brought to the simple form (\ref{nfieldssimple}).  Instead,  if the kinetic term is written as $\frac{1}{2} (\partial\phi_i)^2$, the cosines will depend on nontrivial linear combinations of the  canonical fields $\phi_i$.  In this case we will still define 
$f_i^2$ to be the $i$th eigenvalue of $K_{ij}$, even though  none of the  single cosine terms has period $2\pi f_i$.

\section{Kinetic Alignment}

The fundamental limitation on field displacements  in the theory (\ref{basicL})  comes from the periodicity of the axion potential.
In the absence of monodromy, which we will not invoke, the maximum rectilinear displacement that could be used for inflation is half the diameter
of the fundamental domain in the field space.  Here we have in mind a fine-tuned trajectory moving  simultaneously from the maximum to the minimum of each independent cosine in $V$.  More realistically,  validity of the quadratic expansion around a minimum permits displacements
$\Delta\Phi \sim 0.1 \cdot{\rm{Diam}}$.

Displacements in ${\cal M}_{\theta}$ are related to  meaningful physical displacements in ${\cal M}_{K}$ by the metric on field space, $K_{ij}$.
If  no rotation is required  to relate the lattice basis to the kinetic basis,
the hypercube of side length $2\pi$ in ${\cal M}_{\theta}$ is mapped by the metric information to a rectangular box with side lengths $2\pi f_i$: see figure \ref{fig:illustration}.
The maximum  (rectilinear) displacement of a canonical field in ${\cal M}_{K}$ is then $\pi f_{\rm{Py}}$.

We will be interested instead in the generic situation where the lattice basis and the kinetic basis are related by a nontrivial rotation.
For achieving a large field range, it is advantageous to have the metric assign as much physical distance as possible to a given dimensionless displacement along the particular direction in which ${\cal M}_{\theta}$ has its maximum diameter (namely, $2\pi\sqrt{N}$).
The optimal case is then that the eigenvector $\psi_N$ of $K_{ij}$  with the largest eigenvalue, $f_N^2$, points along a long diagonal, e.g.~along $\frac{1}{\sqrt{N}}(1,1,\ldots 1)$  in the lattice basis.   We refer to this  circumstance as (perfect) {\it{kinetic alignment}}.
In this case the physical diameter of the fundamental region of ${\cal M}_{K}$ is $2\pi\sqrt{N} f_N$.

\section{Mechanisms for Kinetic Alignment}

Although kinetic alignment might appear unlikely at first glance, it is  essentially inevitable in a wide range of systems.
Consider an ensemble of theories of the form (\ref{basicL}); the associated $K_{ij}$ then form an ensemble $\cal{E}$ of $N\times N$ matrices.  Suppose that ${\cal E}$  is statistically rotationally invariant, so that the corresponding normalized eigenvectors $\psi_a$ point in directions  that are uniformly  distributed on $S^{N-1}$.
Then in the large $N$ limit, the components $\psi_{a}^{(i)}$, $i=1,\ldots,N$,
are distributed as
$\psi_{a}^{(i)} \in \frac{1}{\sqrt{N}}{\cal N}(0,1)$, with ${\cal N}(0,1)$ denoting the normal distribution.
Intuitively,  a single component of order unity is possible only if many other components are atypically small.
More geometrically, nearly all eigenvectors point approximately along a diagonal direction in some hyperoctant,  rather than being nearly parallel to a Cartesian basis vector.  This is not surprising, since there are $2^N$ diagonals but just $N$  basis vectors.

This general phenomenon is known as {\it{eigenvector delocalization}}  in random matrix theory, and has been proved to hold  in a number of random matrix ensembles \cite{Erdos09,TaoVu}.
We will focus on the  well-motivated case in which $K_{ij}$ belongs to a canonical ensemble of
positive definite random matrices known as the Wishart ensemble: we take
\begin{equation}
K=A^\top A, \,~~~ \text{with}~ A_{ij}\in \Omega(0,\sigma)\, ,
\end{equation}
where $\Omega(0,\sigma)$ is a statistical distribution with mean zero and variance $\sigma^2$.
Given suitable bounds on the moments of $\Omega(0,\sigma)$, Tao and Vu have proved  that with overwhelming probability, the eigenvectors $\psi_a$ of $A^\top A$ have components
$\psi_a^{(i)} \lesssim {\cal O}(1/\sqrt{N})$, up to corrections logarithmic in $N$ \cite{TaoVu}.
This result is universal, in the sense that it does not depend on the details of $\Omega(0,\sigma)$.
We conclude that  essentially every eigenvector of a Wishart matrix is nearly parallel to a diagonal direction in a hyperoctant.

If the kinetic matrix $K_{ij}$  is well-approximated by a Wishart matrix ---  an assumption that  is natural on the grounds of universality and symmetries, and is substantiated by investigations of random K\"ahler metrics, both in mathematics \cite{Ferrari:2011is} and in string compactifications \cite{Longtoappear} ---  then the eigenvector $\psi_N$ with eigenvalue $f_N^2$  points along a nearly diagonal direction \cite{TaoVu},  along which ${\cal M}_{\theta}$ has diameter $2\pi\sqrt{N}$.  As a result, the diameter of the physical  field space ${\cal M}_{K}$ is
\begin{equation}
{\rm{Diam}} \approx 2\pi f_N \sqrt{N} \ ,  \label{diam2}
\end{equation} where the $\approx$  becomes an equality in the case of perfect alignment.
This is one of our main results.
Let us emphasize that the diameter  of the field space is given by (\ref{diam2}) unless $K_{ij}$ violates eigenvector delocalization: the Wishart model  given as a concrete example is well-motivated and well-understood, but (\ref{diam2}) applies to a much wider range of kinetic terms.

\section{Masses and Misalignment}
Thus far we have established that displacements of order $\sqrt{N}f_N$  are generically possible in systems with $N$ axions.
Whether large-field inflation can occur along such a direction depends on the form of the scalar potential,  which we now consider.

Expanding (\ref{vcos}) to quadratic order around the minimum at $\theta_i=0$, we have $V \approx \frac{1}{2}\Lambda_i^4 \theta_i^2 \equiv \frac{1}{2} M^\text{L}_{ij}\theta^i \theta^j$ with $M^\text{L}={\rm{diag}}(\Lambda_i^4)$.
Rotating and rescaling to canonical fields $\vec{\phi}=\text{diag}(f_i) S^\top_{K}\vec{\theta}$, where
$S_K^\top K S_K = \text{diag}(f_i^2)$,
the Lagrangian takes the form
\begin{equation}
{\cal{L}} = \frac{1}{2} (\partial\phi_i)^2 - \frac{1}{2}M^{\text{C}}_{ij}\phi^i \phi^j\,.
\end{equation}
Diagonalizing $M^\text{C}$ via $S_{M^\text{C}}^\top M^\text{C} S_{M^\text{C}} = \text{diag}(m_i^2)$,
we have
\begin{equation}
{\cal{L}} = \frac{1}{2} (\partial\Phi_i)^2 -\frac{1}{2}  m_i^2 \Phi_i^2 \, ,  \label{Lcan}
\end{equation} with $\vec{\theta}=S_K \text{diag}(1/f_i)S_{M^\text{C}}\vec{\Phi}$.
In the special case that $\Lambda_i = \Lambda~\forall i$, so that $M^\text{L}=\Lambda^4{\mathbb 1}$, the masses
of the  $\Phi_i$ are given by $m_i^2=\Lambda^4/f_i^2$,
while $\vec{\Phi}=\text{diag}(f_i)S_K^\top \vec{\theta}$.
The maximal displacement then corresponds, in this simple case, to motion predominantly by the lightest field.

The extent to which the potential disrupts alignment depends on the precise relation between
$\vec{\theta}$ and $\vec{\Phi}$, and hence on the relation between $K_{ij}$ and $M^\text{L}_{ij}$.
We will begin by presenting an  explicit example with $N=2$, assuming perfect alignment  in the absence of mass terms, and
keeping both the eigenvalues of the kinetic matrix and the diagonal entries of $M^\text{L}$ general: $K=S_K\text{diag}(f_1^2,f_2^2)S_K^\top$ and $M^L=\text{diag}(\Lambda^4_1,\Lambda^4_2)$, where $S_K$ is a rotation by $\pi/4$.
Assuming that $f_1<f_2$ and $\Lambda^4_2>\Lambda^4_1$, and writing $\alpha = f_1^2/f_2^2$, we find that for
\begin{equation}
{8{\alpha }+\sqrt{2}{\alpha^2}-\sqrt{2}\over 1+6 {\alpha }+{\alpha^2}}\le{\Lambda^4_1\over \Lambda^4_2}\le1\,,
\end{equation}
a displacement proportional to $(0,1)$ has a cosine $\gtrsim 0.86$ with the diagonal direction.
This implies that for $f_1/f_2\sim {\mathcal O}(1)$ there is a tight constraint on the hierarchy allowed in the mass matrix in order to preserve alignment.  However, for $f_2/f_1$ sufficiently large, there is no constraint on the hierarchy of $\Lambda^4_1/\Lambda^4_2$.
We conclude that large hierarchies in the $\Lambda_i$ can spoil the alignment mechanism: for some ranges of masses, the dynamically preferred trajectory may not be aligned with a diagonal direction in field space.

In general models with $N \gg 1$ axions, approximate alignment\footnote{A sufficient condition for perfect alignment is that $M^\text{L}$ is diagonalized by the eigenvectors of $K_{ij}^{-1}$.} persists at the level of (\ref{Lcan}) if the hierarchies in the $f_i$  are sufficiently large compared to the hierarchies in the  dynamically-generated scales $\Lambda_i$.  We have checked numerically that for $K_{ij}$ a Wishart matrix, the effective number of fields contributing to inflation equals the number of the  $\Lambda_i^4$ within a factor $\sim 2$  of the smallest of the $\Lambda^4_i$.  While a closely-spaced  spectrum of this form  is not a generic expectation in field theories with $N$  axions,
approximate equality of many $\Lambda_i$  could be achieved through moderately fine-tuned choices of flux in  a string compactification.
We leave a systematic study of this point for the future.

\section{Dynamics and Predictions}

\begin{figure}
  \centering
  \includegraphics[width=.3\textwidth]{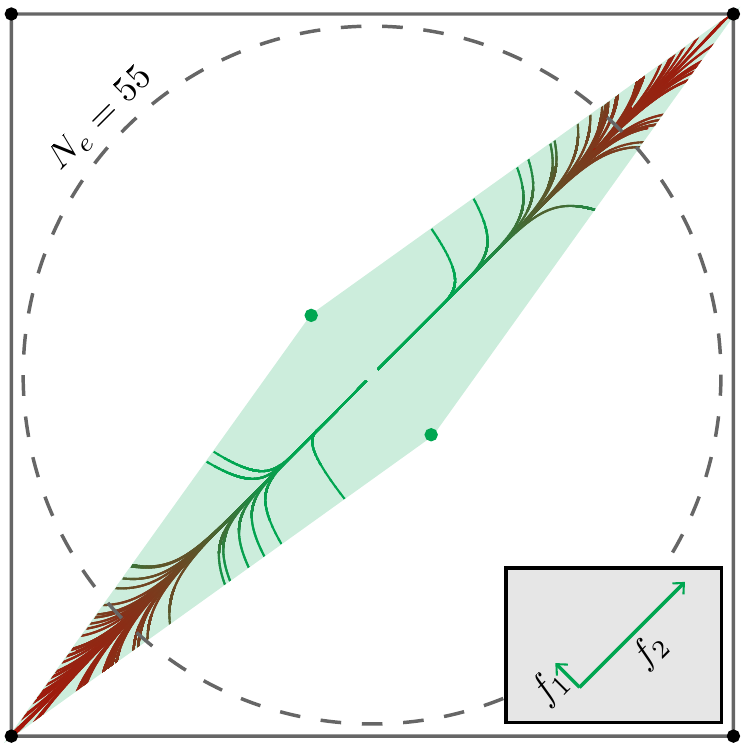}
  \caption{\small Attractor behavior for $N=2$, with axes as in fig.~1.  The  red (green) curves are inflating (non-inflating)  trajectories.  Curves beginning outside the  circle yield $N_{e} \ge 55$.}\label{fig:inflation}
\end{figure}

The dynamics in aligned N-flation  depends on the mass matrix, and in particular on the relative sizes of the $\Lambda_i$.
In the simple case where $\Lambda_i=\Lambda~\forall i$, the field  that admits the largest displacement  is automatically the lightest field.
Inflation arises very naturally in this setting,  for a wide range of initial  conditions: the heavier fields relax toward their minima,  leaving the lightest field, which then drives single-field inflation: see figure \ref{fig:inflation}.
For $f_1,\ldots f_{N-1} \ll f_N$, the heavy fields relax quickly, and the resulting effective description is simply $\frac{1}{2}m^2\phi^2$  chaotic inflation, albeit underpinned by $N$  shift symmetries.
We then have $n_s \approx 0.967$ and $r\approx 0.13$ for $N_e=60$.

More general realizations of aligned N-flation will manifest multi-field behavior.
Provided the slow-roll approximations hold and an adiabatic limit is reached before the end of inflation,
we have \cite{Vernizzi:2006ve,Battefeld:2006sz,Dias:2012nf,Frazer:2013zoa}
\begin{align}
&{\cal P_{\zeta}}= \frac{H^{2}}{4\pi^{2}}N_{e}, \quad && n_{s}-1= -2\epsilon-\frac{1}{N_{e}}, \nonumber \\
&\alpha = -8\epsilon^2-\frac{1}{N_{e}^2}+4\epsilon_{i}\eta_{i}, \quad && r =\frac{8}{N_{e}} \, ,
\end{align}
where we have set $M_P=1$, all quantities are to be evaluated at horizon crossing, $\epsilon \equiv \sum \epsilon_{i}$, $\epsilon_{i}\equiv\frac{1}{2} (V_{i}' /V)^2$, $\eta_i\equiv V_{i}'' /V$ and $3H^2\approx V = \sum V_{i}$, where $V_{i}\equiv \frac{1}{2}m_{i}^{2}\Phi_{i}^{2}$ (no sum),
primes represent differentiation with respect to the corresponding field, and the index $i$ indicates summation over the light fields active during inflation.
When a few fields are active, the power spectrum, spectral index and running  depend on the mass hierarchy and on the initial conditions,
but the single field result still corresponds to an attractor \cite{Frazer:2013zoa}.
The most robust prediction is the tensor-to-scalar ratio, which to a good approximation
depends only on the choice of pivot scale.
Excitations of heavier fields provide the prospect of novel signatures, but we leave a complete study of multifield effects in aligned N-flation  as a subject for future work.
\section{Conclusions}

We considered a theory  containing $N$ axions with decay constants $f_1 \le \ldots \le f_N$, and asked  whether the maximum collective displacement $\Delta\Phi$ can exceed the Pythagorean sum $f_{{\rm{Py}}}$ of the $f_i$ (in the absence of fine-tuning of the relative sizes of the $f_i$, as in \cite{Kim:2004rp,Choi:2014rja,Tye:2014tja}).

The allowed region of field space is defined by the fundamental periodicities  of the dimensionless axions;
without loss of generality, this region can be taken to be a hypercube in $\mathbb{R}^N$.
To convert  dimensionless displacements to physical displacements requires the use of the metric on field space.   If the metric on field space has eigenvectors that align with the edges of the hypercube then the diameter is $2\pi f_{{\rm{Py}}}$, and $\Delta\Phi \approx f_{{\rm{Py}}}$.  But if instead the eigenvector of the metric with largest eigenvalue $f_N^2$  is aligned with a (long) diagonal  of the hypercube, the diameter is $2\pi \sqrt{N} f_{N}$, so that $\Delta\Phi \approx \sqrt{N} f_{N}$, which is considerably larger than $f_{{\rm{Py}}}$  in the generic case in which the $f_i$  are distinct.
We referred  to this situation as {\it{kinetic alignment}}.
Approximate kinetic alignment is  equivalent to the phenomenon of eigenvector delocalization in random matrix theory, which has been proved to hold in a number of relevant cases, in particular if $K_{ij}$  is a Wishart matrix \cite{TaoVu}.   At the level of kinematics, kinetic alignment is almost inevitable  in a system with $N \gg 1$  axions and a general kinetic term.
Although our arguments did not rely on string theory,  and hold in an effective field theory with generic  axion kinetic mixing, the necessary structures are
readily obtained in
compactifications of string theory.

We then argued that  the axion potential (\ref{vcos})  can  be compatible with  large-field inflation along an aligned  direction. As an example, if $K_{ij}$  is a Wishart matrix and
$P \le N$ of the  dynamically-generated scales $\Lambda_i^4$ fall within a range of size $\sim 2$, then $P$  fields participate in the alignment, and the effective range is $\sqrt{P}f_N$.
Alignment is possible  for more general  potentials, but we leave a systematic analysis for the future.

Arranging for $N \gg 1$  axions to have decay constants $f_i$ as large as ${\cal{O}}(0.1) M_P$  is a serious
challenge for the construction of models of N-flation in string theory (cf.~e.g.~\cite{Dimopoulos:2005ac,Easther:2005zr,Kallosh:2007cc,Grimm:2007hs}):
perturbative control of the $g_{\rm{s}}$ and $\alpha^{\prime}$  expansions generally enforces $f_i \ll M_P$ \cite{Banks:2003sx}, and while  accidental cancellations may permit a few of the $f_i$  to be larger, points in moduli space with many $f_i$ large are  not presently computable.   Kinetic alignment allows for successful large-field inflation even if only {\it{one}}  axion has  large (but sub-Planckian)  decay constant,  dramatically reducing the  difficulty of  embedding N-flation  in string theory.

The signatures of aligned N-flation are  very similar to those of single-field $m^2\phi^2$  chaotic inflation \cite{Linde:1983gd}, even though the underlying symmetry structure is distinctive.

\section{Acknowledgements}
We thank Connor Long,  Paul McGuirk, John Stout, and Timm Wrase for very useful conversations on related topics.
The research of T.~C.~B.~and L.~M.~was supported by NSF grant PHY-0757868.
M.~D.~acknowledges support from the European Research Council under the European Unions Seventh Framework Programme (FP/20072013)/ERC Grant Agreement No. [308082].  J.~F.~is supported by IKERBASQUE, the Basque Foundation for Science.

\bibliography{References}

\end{document}